\documentclass{vgtc}                          

\graphicspath{{figures/}{pictures/}{images/}{./}} 

\usepackage{times}                     

\usepackage{tabu}                      
\usepackage{booktabs}                  
\usepackage{lipsum}                    
\usepackage{mwe}                       
\usepackage{cite}
\usepackage{amsmath,amssymb,amsfonts}
\usepackage{algorithmic}
\usepackage{graphicx}
\usepackage{textcomp}
\usepackage{gensymb}
\usepackage{graphicx}
\usepackage{subfigure}
\usepackage{threeparttable}
\usepackage{makecell}
\usepackage{booktabs}
\usepackage{multirow}

\usepackage{mathptmx}                  

\onlineid{9599}

\vgtccategory{Localization, spatial registration, and tracking}

\vgtcinsertpkg

\title{A Monocular SLAM-based Multi-User Positioning System with Image Occlusion in Augmented Reality}

\author{Wei-Hsiang Lien\thanks{e-mail: r10922050@ntu.edu.tw}\\ %
     \scriptsize National Taiwan University, Taipei, Taiwan (ROC) %
\and Benedictus Kent Chandra\thanks{e-mail: r11922178@ntu.edu.tw}\\ %
        \scriptsize National Taiwan University, Taipei, Taiwan (ROC) %
\and Robin Fischer\thanks{e-mail: d08922027@ntu.edu.tw}\\ %
        \scriptsize National Taiwan University, Taipei, Taiwan (ROC) %
\and Ya-Hui Tang\thanks{e-mail: d05922027@ntu.edu.tw}\\ %
        \scriptsize National Taiwan University, Taipei, Taiwan (ROC) %
\and Shiann-Jang Wang\thanks{e-mail: lerdahwang@jorjin.com.tw}\\ %
     \scriptsize JorJin Technologies Inc., New Taipei, Taiwan (ROC) %
\and Wei-En Hsu\thanks{e-mail: vincenthsu@jorjin.com}\\ %
     \scriptsize JorJin Technologies Inc., New Taipei, Taiwan (ROC) %
\and Li-Chen Fu\thanks{e-mail: lichen@ntu.edu.tw}\\ %
     \parbox{1.4in}{\scriptsize National Taiwan University, Taipei, Taiwan (ROC)}}

\abstract{
In recent years, with the rapid development of augmented reality (AR) technology, there is an increasing demand for multi-user collaborative experiences. Unlike for single-user experiences, ensuring the spatial localization of every user and maintaining synchronization and consistency of positioning and orientation across multiple users is a significant challenge. In this paper, we propose a multi-user localization system based on ORB-SLAM2 using monocular RGB images as a development platform based on the Unity 3D game engine. This system not only performs user localization but also places a common virtual object on a planar surface (such as table) in the environment so that every user holds a proper perspective view of the object. These generated virtual objects serve as reference points for multi-user position synchronization. The positioning information is passed among every user's AR devices via a central server, based on which the relative position and movement of other users in the space of a specific user are presented via virtual avatars all with respect to these virtual objects. In addition, we use deep learning techniques to estimate the depth map of an image from a single RGB image to solve occlusion problems in AR applications, making virtual objects appear more natural in AR scenes.} 


\begin{document}



\maketitle
\section{Introduction} 
Augmented Reality technology superimposes 3D virtual models, images, and sound onto the real world, allowing people to see and interact with virtual objects, using devices such as smartphones, tablets, and AR glasses. With the rapid development of Augmented Reality(AR) technology over the past few years, more and more AR applications are moving towards a multi-user experience. Regardless of the domain, AR applications typically require the implementation of technologies such as user positioning and tracking, environmental scene understanding, and occlusion handling.

 Multi-user positioning and tracking is a critical technology in AR applications. Its main purpose is to determine the user's position and orientation in the real world and transmit each user's position information to other users' devices using network communication technology to achieve accurate placement of virtual avatars. Positioning technology typically includes marker-based and markerless positioning methods. 

Marker-based AR positioning requires placing specific real-world markers, such as QR codes or images, and using cameras to identify and track them. This method can achieve high positioning accuracy because the markers provide clear position and orientation information, making it easy to place virtual objects in precise locations. We can easily calculate and synchronize the position and orientation of each user based on the marker. However, if there are no markers within the user's field of view, the tracking functionality cannot continue. On the other hand, Markerless AR positioning technology can determine virtual object placement by identifying feature points in the scene without requiring special markers, making it more convenient. This method captures feature points in the real scene, such as walls, tables and chairs, using cameras or other sensors, and uses computer vision techniques for identification and tracking. Markerless positioning such as ORB-SLAM2\cite{b1} use feature points in the environment to determine the user's camera pose. However, there are currently only few mature methods\cite{Stotko_2019, DBLP:journals/corr/abs-2108-08325} for aligning the multiple coordinate systems of multiple users respectively staying in different locations using this technique.

Environmental scene understanding technology is one of the most important techniques in AR applications. This technology is typically used to determine appropriate positions at which the virtual objects should be placed, such as the surfaces of tables, in the real world. Occlusion is also an important issue that needs to be addressed in AR applications. Occlusion refers to the accurate representation of virtual objects being blocked or hidden by real-world objects. In traditional AR applications, virtual objects are superimposed directly onto the image, which makes the virtual objects persistently appear on top of and block the real objects. This can significantly impact the realism and interactivity of the AR application.

 Therefore, in this work, we propose a novel AR system that addresses the challenge of synchronizing the localization of multiple users staying in different spaces to pursue an AR experience such that users will view all other's avatars in their physical spaces and by determining appropriate planes for placing virtual object to be viewed by all users also for another AR experience. Additionally, the system aims to estimate the depth map of the environment from a single RGB image, specifically designed for occlusion handling.

\section{Related Works}

\subsection{Simultaneous Localization and Mapping}

SLAM is the abbreviation for Simultaneous Localization and Mapping. Localization aims to estimate the sensor's position and orientation in the environment. Mapping, focuses on constructing a representation or model of the environment. 

        One of the well-known works in the field of visual SLAM is ORB-SLAM\cite{b2} (Oriented FAST and Rotated BRIEF SLAM). It utilizes the ORB feature descriptors for robust and efficient feature matching. ORB-SLAM operates by extracting ORB features from camera frames and tracking them over time to estimate the camera's motion and build a sparse 3D map of the environment. It incorporates bundle adjustment techniques to refine the map and camera poses, improving the accuracy of the estimated trajectory. ORB-SLAM2\cite{b1} is an extension of ORB-SLAM that introduces additional features such as multi-threading, keyframe database management, and loop-closing methods. We decided to utilize their solution as our baseline for real-time visual SLAM, especially since it's open to different platforms, including mobile devices.
        Other popular SLAM solutions include Vuforia \cite{b11} and OS based providers by Google and Apple, which are all closed source and requires developers to purchase license. These solutions being closed source also make developers unable to modify the algorithm to suit our need. However, we still analyzed these solutions for accuracy comparison.

\subsection{Multi-user AR Collaboration}
Currently, numerous augmented reality applications incorporate SLAM-based positioning technology to track user localization. However, traditional SLAM frameworks are designed for single-user operation, lacking support for map sharing between clients (\cite{b3, b4}). Some frameworks involve multiple collaborating robots, where clients create local maps that are merged remotely, yet this approach often incurs high communication costs due to the transmission of large mappoints and keyframes (\cite{b5, b6}).

Recent studies have explored novel approaches to enable multi-user AR collaboration (\cite{b7, b8, b9, b10, 9089441, DBLP:journals/corr/abs-2108-08325, wang_scene_2023}). In these systems, each AR device independently performs SLAM to capture visual features of the physical space relative to its local coordinate system. Users then share these visual maps to establish a common coordinate system and estimate the positions of other users. For instance, SLAM-share utilizes an edge-centric architecture to centrally combine and distribute maps to all users (\cite{b10}). Conversely, Apple ARKit employs a peer-to-peer architecture, where the host of the AR session shares its current map with joining users.

However, a significant challenge arises when merging local maps into a global map, particularly in different physical spaces where there are no shared feature points for alignment. This reliance on shared visual features limits the applicability of these systems and makes achieving accurate alignment challenging.

In summary, while current multi-user AR collaboration approaches show promise, addressing the challenge of map fusion across different physical spaces remains an area for further research and innovation.

\section{Methodology}
In this paper, we present a novel AR system which can perform multi-user positioning for multiple users staying in different places aimed at enhancing their AR collaboration experiences by enabling them to interact within a shared augmented environment. The AR devices need in our proposed system are Jorjin J7EF Plus AR glass\cite{b17}, each of which is equipped with an RGB camera for real-time image capture of their surroundings. These captured images serve as input for various functionalities, including motion tracking, plane estimation, and occlusion. We will first provide an overview of our system and subsequently explain the algorithmic details and implementation specifics of each module.
\begin{figure}[tb]\centering
    \includegraphics[scale=0.32]{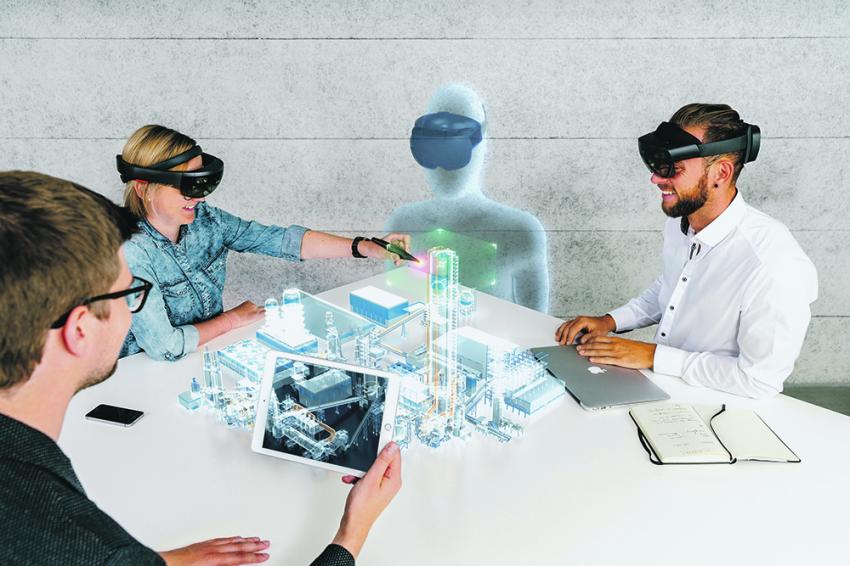}
    \caption{Work collaboratively in augmented reality\cite{b21}.}\label{fig:remote_collaborate}
\end{figure}

\subsection{System Overview}

\begin{figure*}[tb]\centering
            \includegraphics[width=\textwidth]{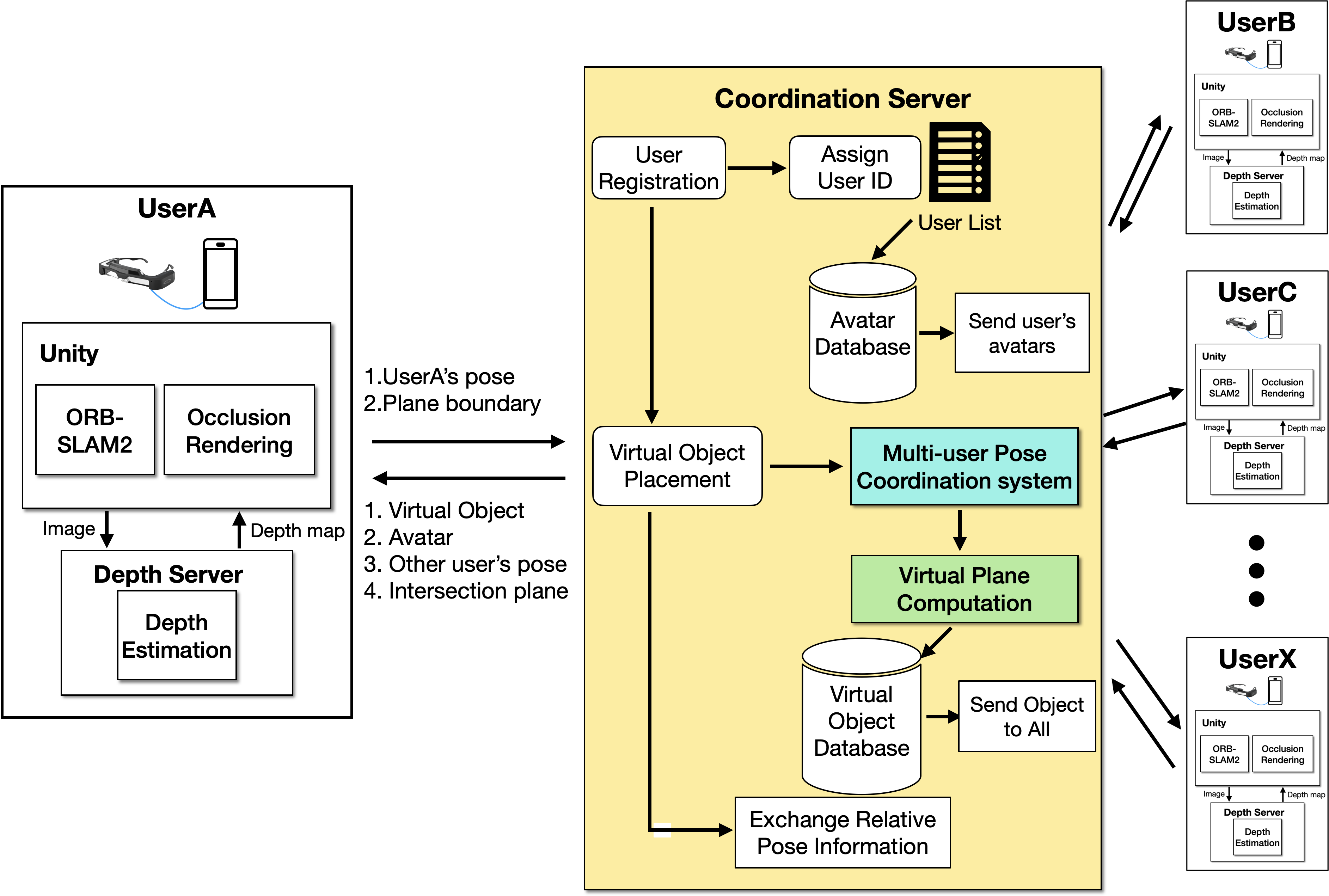}
            \caption{System overview of the multi-user positioning system.}\label{fig:system_overview}
    \end{figure*}

 Fig.~\ref{fig:system_overview} provides the architecture of our designed multi-user positioning system, with the conceptual target application shown in Fig.~\ref{fig:remote_collaborate}. A group of people wearing AR glasses are sitting around a work table having a discussion, with one of the virtual avatars as a remote participant. This allows participants to communicate across physical boundaries and see virtual objects and information in their environment. Our proposed system can be decomposed into four major parts. 
 \begin{enumerate}
     \item The localization module, which utilizes RGB images captured by the camera, through the employment of the SLAM algorithm, it accurately estimates the user's camera pose which is used to update the virtual camera in Unity, enabling the correct rendering of virtual objects from the user's perspective.
     \item The plane estimation module, which utilizes the mapping information generated by SLAM process to identify suitable planes within the environment for placing virtual objects. 
     \item The coordination server, which coordinates the coordinate systems and the plane's information among multiple users, facilitating information exchange and collaboration among all participants.
     \item The depth server, which employs a deep-learning model to estimate depth maps of real environment for each frame, and handle the occlusion problem with these depth map in Unity, enhancing the overall realism of the AR experience.
 \end{enumerate}  

\subsection{Localization Module}
The proposed localization module utilizes the open-source ORB-SLAM2\cite{b1}, which offers real-time 6 DOF tracking and simultaneous mapping of the surrounding environment. It is important to note that the open-source ORB-SLAM2 library is implemented in C++, while Unity is a C\#-based game engine. Therefore, the SLAM library cannot be directly used within Unity. To integrate the SLAM algorithm into the Unity Engine, we rely on the native plugins offered by Unity. This involves transforming the ORB-SLAM2 system into a dynamic or shared library, which can then be accessed by the Unity Engine as a plugin. By leveraging these native plugins, we enable Unity to call the required SLAM functions and incorporate them into our application. 

\begin{figure}[tb]\centering
            \includegraphics[scale=0.13]{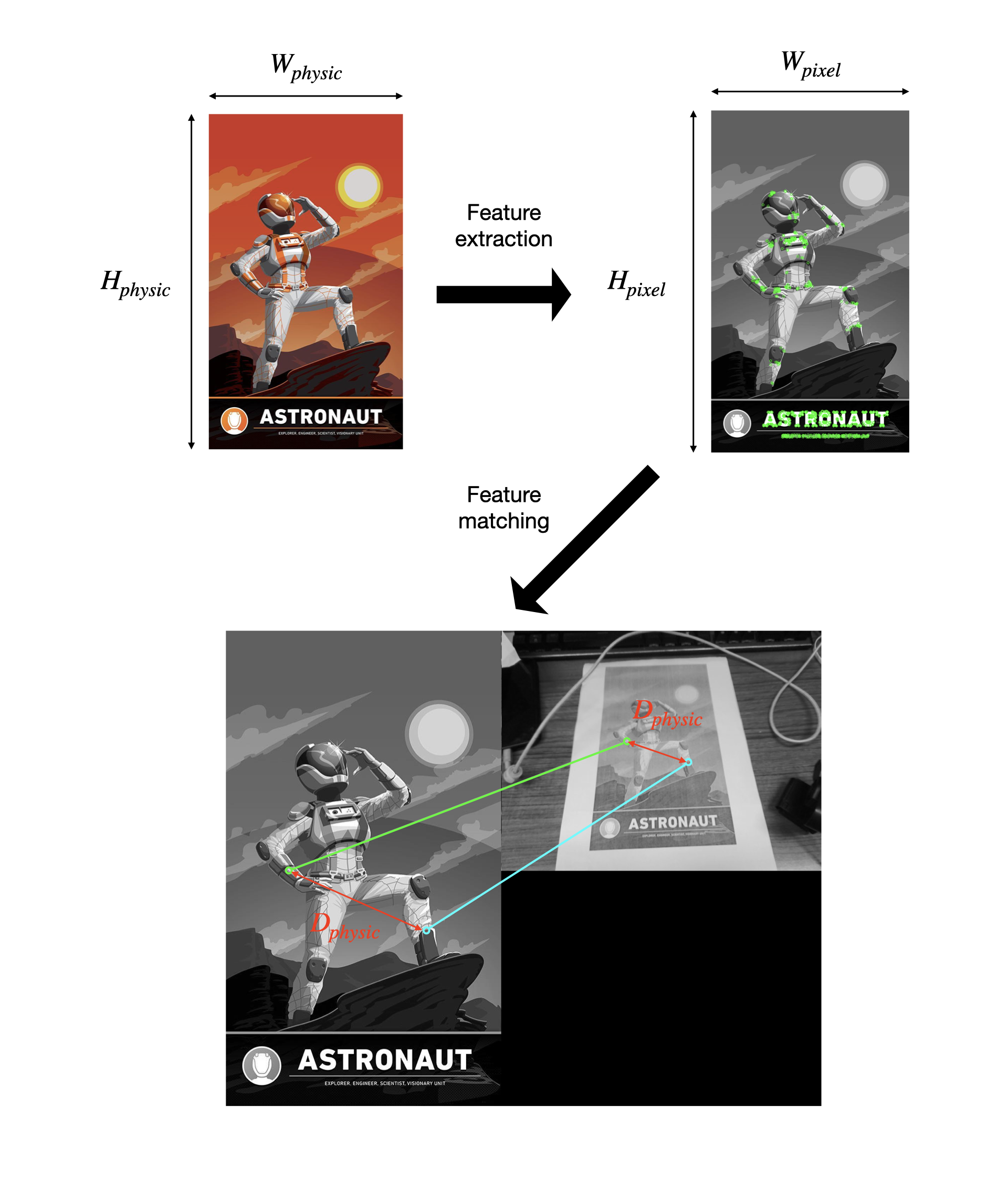}
            \caption{The process of scale calibration.}\label{fig:scale}
    \end{figure}

In our localization module, we utilize a monocular SLAM approach, which relies solely on an RGB camera configuration. One limitation of monocular SLAM is the existence of scale uncertainty, leading to potential scale drift or inconsistency in the reconstructed environment. To address the scale uncertainty inherent in monocular SLAM, we have implemented a calibration step using a calibration marker with precisely measured size. The process of scale calibration is illustrated in Fig.~\ref{fig:scale}.
 Initially, we utilize a marker provided by Vuforia\cite{b11}, which has dimensions of width $W_{pixel}$ and height $H_{pixel}$ in terms of pixels. We then measure its physical size in the real-world environment, where the marker's width and height are denoted as $W_{physic}$ and $H_{physic}$, respectively. It is important to note that this marker can be replaced with any image containing sufficiently distinct and recognizable feature points. And then, we extract the ORB features from the marker to establish correspondences with the scene image captured by the camera. This is achieved through ORB feature matching, where we compare the extracted ORB features from the marker with those obtained from the scene image. ORB features of the scene image are detected using the SLAM algorithm. By comparing these features, we can determine the corresponding keypoints between the marker and the scene image. After identifying the correspondences between the marker and the scene image, we specifically focus on selecting the two correspondences with the highest similarity. With these chosen correspondences, we can employ the following formula to calculate the physical distance $D_{physic}$ between these two feature points:
    \begin{equation}
        D_{physic} = \frac{{W_{physic}}}{{W_{pixel}}} \times {||\bf{p_1}-\bf{p_2}||}
    \end{equation}
    where $\bf{p_1}$ and $\bf{p_2}$ are the correspondences in the marker. $D_{physic}$ also represents the physical distance between correspondences in the scene image. We further utilize the 3D map points generated by the SLAM algorithm for these two correspondences in the real scene image. By utilizing both the physical distance $D_{physic}$ and the distance between the 3D map points, which represents the SLAM distance, we can obtain the scale factor by calculating the ratio between the physical distance and the SLAM distance using the following formula:
    \begin{equation}
        Scale = \frac{D_{physic}}{||\bf{P_1}-\bf{P_2}||}
    \end{equation}
    where $\bf{P_1}$ and $\bf{P_2}$ are the 3D map points of the two correspondences.

    By performing scale calibration, we can achieve a seamless alignment of the coordinate systems between the SLAM map and the real-world environment. This process enables an accurate representation of virtual elements within the physical surroundings, ensuring precise spatial relationships and measurements. Note that the scale calibration process only needs to be executed once, after which our system maintains the localization without a marker until the application is closed.

\subsection{Plane Estimation}
In AR collaboration scenarios, participants are typically staying in their own environments, and likely working on some virtual objects placed on a table nearby. Conceivably, it is essential for every user to identify the surface of the table in his/her space and place the virtual objects on it. Since the size of the table of every user may be varying, it becomes crucial to determine the intersection area of all detected table planes where virtual objects can be placed reasonably and be interacted with users in different spaces. To estimate planes in the environment, we utilize the map generated by the SLAM process. Assuming that the table plane is not featureless, we extract the 3D map points from the current frame, we can identify the plane that best fits this point cloud. However, the point cloud may contain both inlier points belonging to the desired plane and outlier points that do not lie on the plane. These outliers can adversely impact the accuracy of our plane estimation. Therefore, to address this issue, we employ the RANSAC (RANdom SAmple Consensus) algorithm \cite{b12} to effectively remove outliers before proceeding with the plane estimation. 

After removing the outliers using the RANSAC algorithm, we can enhance the accuracy of our plane estimation by performing Least Square fitting to the remaining inlier points. The optimal plane center position is determined as the centroid of all the inlier points. The normal vector of a plane is found by identifying the singular vector through the utilization of Singular Value Decomposition (SVD) for solving the Least Squares fitting problem. Furthermore, we determine the plane's pose by considering the y-axis as the normal vector. The x-axis direction is calculated by projecting the vector from the plane's origin to the camera's position onto the plane. This allows us to orient virtual objects to face the users. Lastly, the z-axis direction is determined by taking the cross product of the x-axis and y-axis vectors. With the position and orientation of the plane, we can establish the plane's pose and use this information to place the virtual objects in the environment.

However, the plane we estimated is an infinite plane, which means it does not have explicit boundaries. In an AR collaborative scenario, it's advantageous to constrain the dimensions of virtual objects within the bounds of the tabletop area. Therefore, it's important to determine the boundary information of the plane. To address this problem, we leverage the inliers identified by the RANSAC algorithm to determine the boundary of the plane. In the SLAM system, we can obtain the keypoints of these inliers, which refer to the coordinates of the inlier points in the 2D image, as well as their corresponding map points, which represent the three-dimensional coordinates of the inliers in the SLAM map. After obtaining the keypoints and their corresponding map points for the inliers within the SLAM system, the next step is to find the convex hull among these points using the Graham algorithm \cite{b13}. The convex hull of a set of points is the smallest convex polygon that encompasses all the points within the set. Therefore, we can consider it as the boundary for the set.

\subsection{Coordination Server}

\begin{figure*}[tb]
    \centering
    \includegraphics[width=\textwidth]{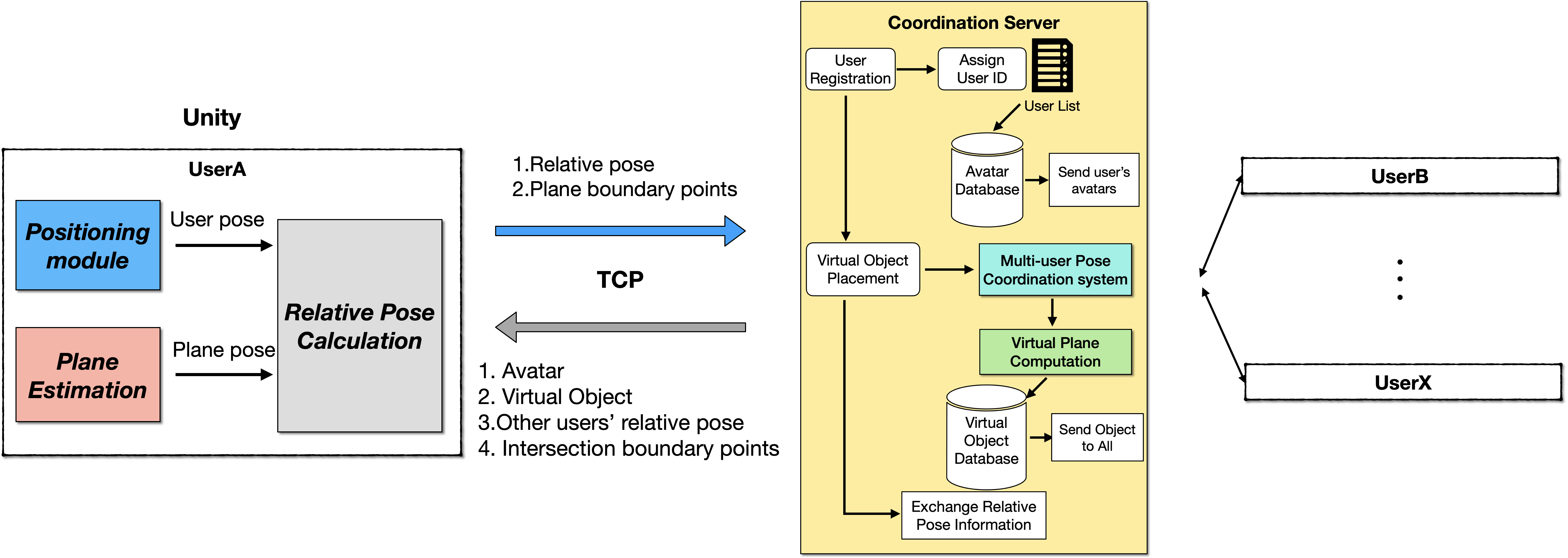}
    \caption{Client-server framework between Unity and coordination server.}
    \label{fig:sync_server}
\end{figure*}

In a multi-user collaborative augmented reality system, the exchange of information between users is essential for facilitating interaction and coordination within the shared virtual environment. We design a coordination server using the TCP protocol to exchange camera pose and plane information among users, where the server-client framework is shown in Fig.~\ref{fig:sync_server}. Camera's position and orientation data enable users to perceive their collaborators' relative poses, whereas the sharing of plane information allows users to establish a shared understanding of the physical surfaces available for virtual object placement or interaction.

\subsubsection{Relative Pose Calculation}

ORB-SLAM2 generates a unique coordinate system for each user, where origin is defined based on the initial camera pose estimation. To coordinate the poses of different users in a shared virtual environment, directly using the camera poses estimated by SLAM is not feasible. An alternative approach is adopted by using a common reference object that exists in every user's environment. In our collaborative scenario, a table is chosen as the reference object, as it is present and necessary for every user to work with. 

Both UserA and UserB should have a physical table in their environment. By utilizing the plane information, we can align the coordinate system among every user. Additionally, we have access to camera pose information from the SLAM's coordinate system, allowing us to determine the relative pose between the user and the plane. These relative poses can be exchanged between users, meaning that the relative pose between UserA and the plane in his/her environment can be transmitted to UserB. UserB can then apply this relative pose to the plane coordinate system in his environment, enabling the calculation of the position and orientation of UserA's avatar.

To achieve this, we begin by utilizing the pose of the plane obtained by the plane estimation module to establish the plane's coordinate system for each user. For instance, for UserA, the plane's pose with respect to the UserA's SLAM coordinate system $\boldsymbol{\mathit{S_A}}$ is denoted as $\boldsymbol{\mathit{_{P}^{S_A}T}}$. Additionally, we obtain the UserA's camera pose with respect to the SLAM coordinate system from the localization module, which is denoted as $\boldsymbol{\mathit{_{C_A}^{S_A}T}}$. The transformation matrix $\boldsymbol{\mathit{_{S_A}^{P}T}}$ that transform the SLAM coordinate system to the plane coordinate system can be calculated by the following formula:
\begin{equation}
        {\boldsymbol{\mathit{_{S_A}^{P}T}}} = \boldsymbol{\mathit{_{P}^{S_A}T}}^{-1}
\end{equation}
To calculate the relative pose of UserA's camera with respect to the plane, we can transform the camera pose from the SLAM coordinate system to the plane coordinate system. This transformation can be done using the following equation:
\begin{equation}
        {\boldsymbol{\mathit{_{C_A}^{P}T}}} = {\boldsymbol{\mathit{_{S_A}^{P}T}}} \cdot {\boldsymbol{\mathit{_{C_A}^{S_A}T}}}
\end{equation}
where $\boldsymbol{\mathit{_{C_A}^{P}T}}$ represents the relative pose of UserA's camera with respect to the plane. $\boldsymbol{\mathit{_{S_A}^{P}T}}$ is the transformation matrix that converts coordinates from the SLAM coordinate system to the plane coordinate system, and $\boldsymbol{\mathit{_{C_A}^{S_A}T}}$ denotes the camera pose with respect to UserA's SLAM coordinate system.

To enable the recovery of UserA's position and orientation with respect to other users' SLAM coordinate systems, the relative pose calculated for UserA's camera with respect to the coordinate system of the plane is transmitted to all other users' devices. This transmission facilitates the consistent alignment of reference frames among users. Each user treats the plane coordinates, established by the plane estimation module, as the shared reference frame. By applying the received relative pose information, other users can transform UserA's position and orientation from the plane coordinate system to their own SLAM coordinate systems using the following equation:
\begin{equation}
        {\boldsymbol{\mathit{_{C_A}^{S_B}T}}} = {\boldsymbol{\mathit{_{P}^{S_B}T}}} \cdot {\boldsymbol{\mathit{_{C_A}^{P}T}}}
\end{equation}
where $\boldsymbol{\mathit{_{C_A}^{S_B}T}}$ represents the relative pose of UserA's camera with respect to the UserB's SLAM coordinate system $\boldsymbol{\mathit{S_B}}$, $\boldsymbol{\mathit{_{P}^{S_B}T}}$ is the UserB's plane coordinate system with respect to the UserB's SLAM coordinate system, and $\boldsymbol{\mathit{_{C_A}^{P}T}}$ denotes the relative pose of UserA's camera with respect to UserA's plane coordinate system. Note that here we coincide all the plane's coordinate systems for convenience of subsequent design of user coordination.

\subsubsection{Multi-user Pose Coordination System}

We have designed two interaction modes. In the "Classroom Mode" we assume that all users are initially positioned on the same side of the virtual objects for discussion and interaction. To achieve this, during the plane estimation process, we define the x-axis of the plane to point towards the users. Since we placed virtual objects facing along the x-axis in Unity, we can ensure that all users are positioned in front of the virtual object. Additionally, leveraging the relative pose calculation mentioned earlier, each user and their corresponding avatar will also be positioned on the same side of the virtual object.

\begin{figure}[tb]\centering
        \includegraphics[width=\columnwidth]{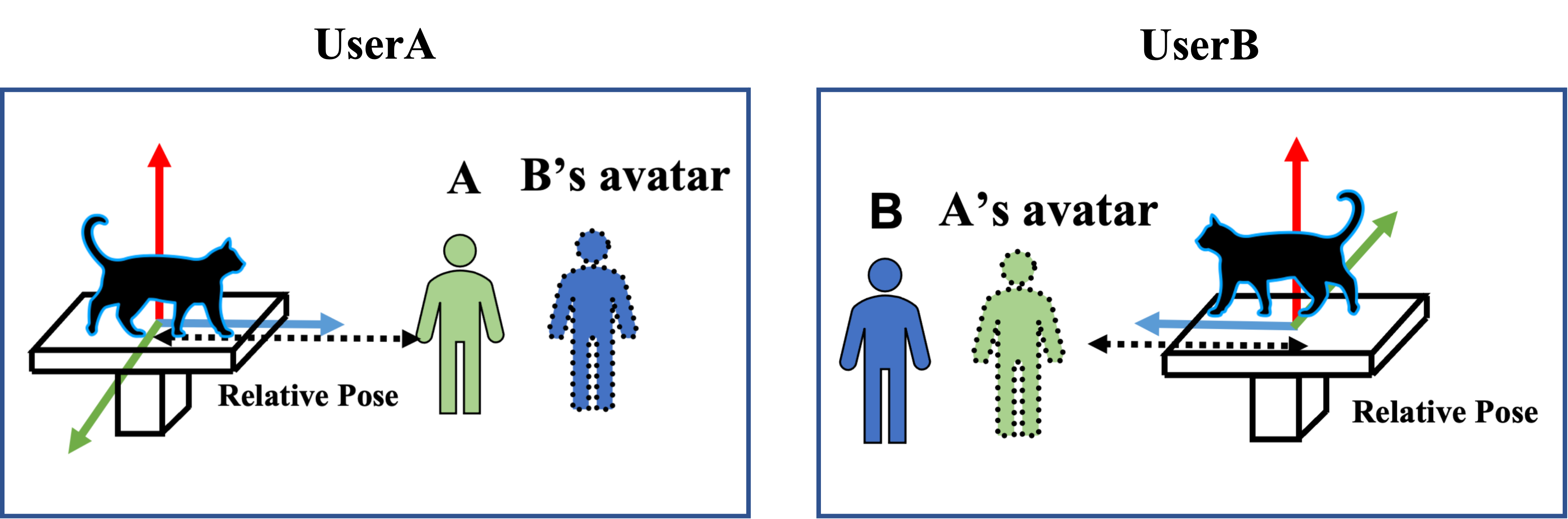}
        \caption{Classroom Mode: Two Users are facing the virtual cat on the table.}\label{fig:classroom_coordinate}
\end{figure}

\begin{figure}[tb]\centering
        \includegraphics[width=\columnwidth]{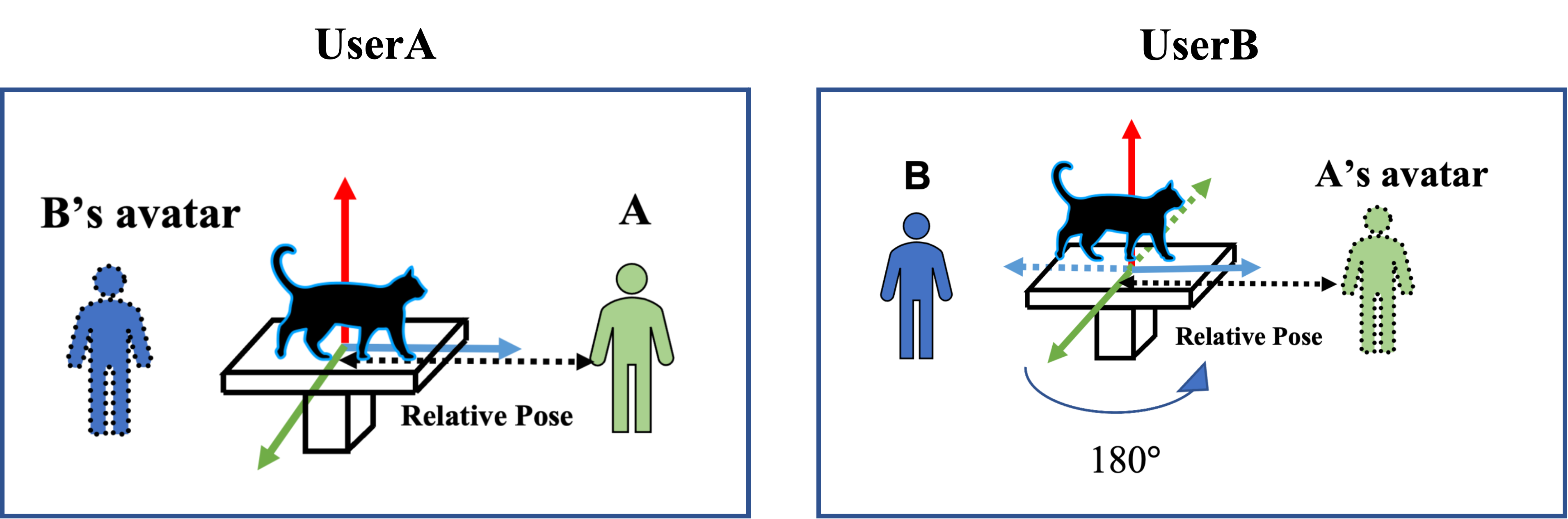}
        \caption{Collaboration Mode: After rotating UserB's plane coordinate system, two users are on the opposite side of the table.}\label{fig:collaborative_coordinate}
\end{figure}

    The second mode we have designed is called "Collaboration Mode". In Collaboration Mode, we assume that every collaborator works around a table instead of being positioned on the same side. Therefore, all of the users should be positioned around the table. For instance, if there are two users, they should be placed on opposite sides of the table, creating a 180-degree angle between two users. To achieve this positioning, we assign a unique ID number to each user, ranging from 0 to $N-1$, and we calculate the angle that needs to be allocated for adjacent two users by the following formula:
    \begin{equation}
            \theta = 360 \degree / N \times i
    \end{equation}
    where $N$ is the total number of the participants and $i$ is the ID number of the user. Once the angle is determined, we apply a rotation transformation to the coordinate system of the plane along its y-axis. This rotation adjusts the relative pose between that specific user and the plane, effectively positioning every user around the table in the initial phase. As shown in Fig.~\ref{fig:classroom_coordinate}, both UserA and UserB are facing the virtual cat on the table in Classroom mode. In Collaboration mode, with a total of two participants, we rotate UserB's plane coordinate system by 180 degrees. After calculating the relative pose, UserA and UserB are positioned on opposite sides of the table, as illustrated in Fig.~\ref{fig:collaborative_coordinate}.


\subsubsection{Virtual Plane Computation}
In the remote interactive scenario, the sizes of tables in each user's environment may differ. This can lead to a situation where a user with a larger table moves a virtual object beyond the boundaries of a user with a smaller table. Therefore, we need to determine the common usable plane area among these users' planes, which represents the intersection of these planes. In the previous section, we employ the Graham algorithm\cite{b13} to find the convex hull of the planes and convert the corresponding SLAM map points within the convex hull to the plane coordinate system representation. After the conversion, the coordinates of these boundary points are in the form (x, 0, z) (since we align the normal vector of the plane with the y-axis). We then transmit these boundary points to the coordination server using the TCP protocol. By extracting the x and z coordinates from the boundary points, we form a set of 2D points on the xz plane. Our objective is to analyze these points on the xz plane in order to identify the intersecting region of all users' planes.

    The intersection area of two planes can be determined using the following properties regarding the intersection of convex polygons\cite{b14}:
    
    \begin{enumerate}
            \item Any vertex from one polygon that lies within the other polygon is also a vertex of the intersection shape.
            \item The points where the edges of two polygons intersect are vertices of the intersection shape.
            \item Any edge from one polygon that lies entirely within the other polygon is also an edge of the intersection shape.
            \item The intersection of two convex polygons forms a convex polygon.
    \end{enumerate}

By applying the properties mentioned, we can determine the vertices and intersection points that constitute the intersection polygon. Firstly, we apply the first property to identify the vertices from each polygon that are contained within the other polygon. In this example, vertices A, C, and D satisfy this condition. Next, we utilize the second property to compute the intersection points between the edges of the two polygons. In the given example, intersection points B and E are obtained. By considering these vertices and intersection points, we can construct the convex hull, as the fourth property states that the intersection of two convex polygons is also a convex polygon. After obtaining the intersection convex polygon, we can use it iteratively to find the intersection with each user's plane. Once the computation is complete, the boundary points of the final intersection area are transmitted to the AR devices of all users, enabling them to interact within the shared intersection area.

\subsection{Occlusion Rendering Module}
In the previous sections, we derived the camera pose using the localization module, acquired the plane pose through the plane estimation module, and received the relative pose of other users from the coordination server. This information is then employed in Unity to position and orientate virtual content. By utilizing the camera pose, we adjust the virtual camera's position and angle, ensuring the correct visualization of virtual content. The plane pose enables us to determine the placement and orientation of virtual objects on the table. Additionally, the relative pose of other users allows us to update the corresponding avatar's pose accordingly. After placing virtual objects in the virtual environment, Unity renders them on the screen. However, similar to traditional AR applications, the virtual content always appears on the top layer of the image layout, blocking the real objects regardless of the virtual content's distance from the camera.

To address the occlusion problem in augmented reality, we adopt the approach proposed in MiDaS\cite{b15} to predict depth from a single image. The authors propose a mixing strategy to blend samples from different datasets, effectively addressing issues such as dataset bias and domain shift that commonly arise when applying depth estimation models to unseen datasets. By employing this approach, the model can achieve enhanced generalization capabilities and adaptability to various scenarios. 

\begin{figure}[tb]\centering
            \includegraphics[width=\columnwidth]{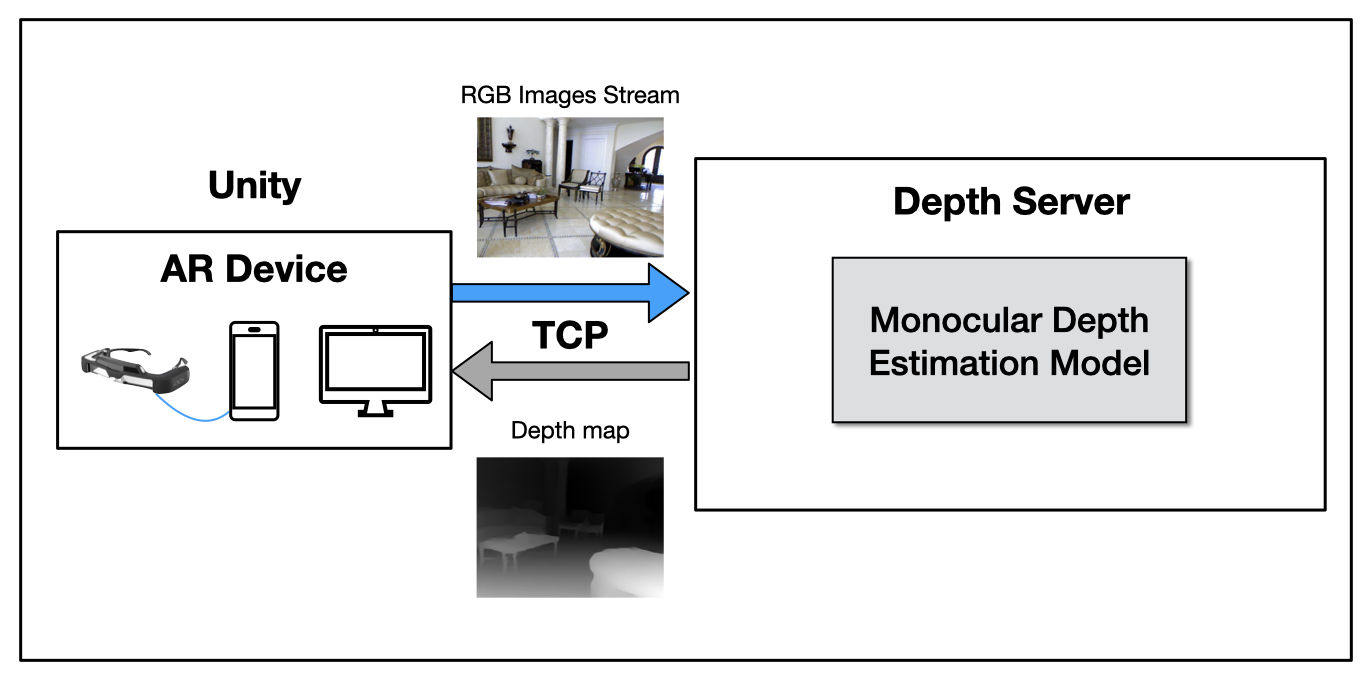}
            \caption{The server-client architecture for monocular depth estimation.}\label{fig:occlusion}
    \end{figure}

Due to the significant computational resource and memory requirements of deep learning models, running them directly on AR devices can lead to performance degradation and increased battery consumption. Therefore, we design a server-client architecture where the deep learning models are executed on the server, and the prediction results are transmitted back to the Unity client. The server-client architecture is shown in Fig.~\ref{fig:occlusion}.

In the Unity environment, we establish a TCP connection between the AR device and the depth server. The captured images are transmitted to a depth server through this connection. The depth server runs the monocular depth estimation model on the received images and generates the gray-scale depth maps. Once the depth estimation process is complete, the depth server transmits the generated depth map back to the AR device through the TCP connection. 

After receiving the depth map, we can employ post-processing techniques in Unity to render virtual objects with occlusion effects. To accomplish this, we utilize the fragment shader, which enables us to modify the appearance of individual pixels during the rendering process. By comparing the depth values from the depth map with the depth of virtual objects in the Unity virtual scene, we can determine which pixels should be rendered with the color of the background image from the real-world scene and which pixels should be rendered with the color of the virtual objects.
\section{Experiments}

\subsection{Experimental Setup}
Our system consists of three main components: AR devices, a coordination server, and a depth server. The AR devices used in our system include Samsung Galaxy S21 smartphones and Jorjin J-Reality J7EF PLUS AR glasses\cite{b17}. The coordination server, responsible for facilitating information exchange among all participants, operates on the Windows 10 operating system. The depth server, hosting the monocular depth estimation model, is built on the Ubuntu 18.04 operating system and is equipped with an RTX 2080Ti GPU.

\subsection{Scale Calibration Results}
 We utilize a known size marker to calibrate the scale of our SLAM system. To evaluate the accuracy of the scale calibration in our positioning system, we measure the relative pose error (RPE) of the camera pose estimated by our positioning module before and after scaling. We use the camera pose output provided by Vuforia as the ground truth for reference. During the experiment, we move the AR device around the table and record the camera poses using both methods. In this setting, we can be considered the camera pose provided by our system as the avatar pose within the Vuforia application, and the camera pose provided by Vuforia can be considered as the ground truth of the avatar pose.

\begin{figure}[tb]\centering
        \includegraphics[width=0.5\columnwidth]{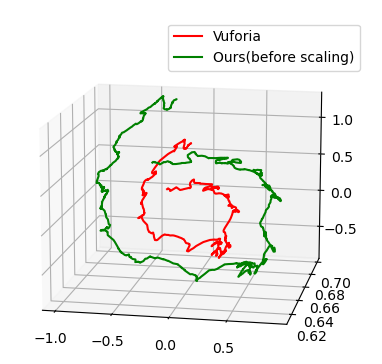}
        \caption{The estimated pose trajectory of our system before scaling compared to Vuforia.}\label{fig:trajectory_a}
\end{figure}

\begin{figure}[tb]\centering
        \includegraphics[width=0.5\columnwidth]{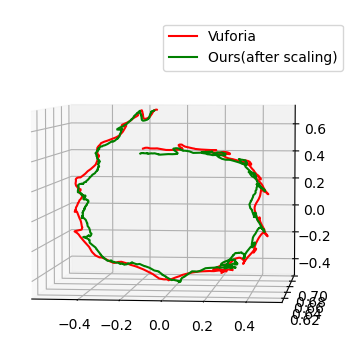}
        \caption{The estimated pose trajectory of our system after scaling compared to Vuforia.}\label{fig:trajectory_b}
\end{figure}

Fig.~\ref{fig:trajectory_a} and Fig.~\ref{fig:trajectory_b} illustrates the trajectories of the camera pose before and after scaling, using Vuforia's trajectory as the ground truth. From Fig.~\ref{fig:trajectory_a}, we can observe that prior to calibrating the scale, although the shape of the trajectory estimated by our positioning system is similar to Vuforia's trajectory, there is a noticeable scale discrepancy. In Fig.~\ref{fig:trajectory_b}, the trajectory of the camera pose after scale calibration aligns more closely with the ground truth. To evaluate the effectiveness of our scale calibration process, we calculate the translational and rotational errors of the relative pose error (RPE). The results of the quantitative comparison of the relative pose error before and after the scale calibration process are presented in Table~\ref{Tab:quan_scale1} and Table~\ref{Tab:quan_scale2}. The results indicate that the camera's rotational error remains consistent irrespective of scaling. Both before and after scale calibration, the Root Mean Square Error (RMSE) of the rotational error is observed to be 0.6302 degrees, and the maximum rotational error is measured at 3.9706 degrees. However, the translational error exhibits significant improvement after scaling. Prior to scaling, the RMSE of the translational error is measured at 0.0112 meters, with a maximum drift of 0.0582 meters. After scale calibration, these values are improved to an RMSE of 0.0065 meters and a maximum drift of 0.0294 meters. Overall, these quantitative results indicate that our scale calibration process effectively reduces the translational error and improves the accuracy of our positioning system.


\begin{table}[tb]
\centering
\renewcommand\arraystretch{1.5}
\caption{The translational error before and after scaling}
\label{Tab:quan_scale1}
\begin{threeparttable}

\begin{tabular}{ |c|c|c|}
\hline
 & \makecell[c]{Translational Error \\ (before scaling) } &
 \makecell[c]{Translational Error \\ (after scaling) } \\
\hline
RMSE & 0.0112m & 0.0065m \\
\hline
Mean & 0.0077m & 0.0049m \\
\hline
Median & 0.0051m & 0.0037m \\
\hline
Max & 0.0582m & 0.0294m \\
\hline
S.D. & 0.0081m & 0.0043m \\
\hline
\end{tabular}
\begin{tablenotes}
\footnotesize
\item S.D.: Standard Deviation
\end{tablenotes}
\end{threeparttable}
\end{table}

\begin{table}[tb]
\centering
\renewcommand\arraystretch{1.5}
\caption{The rotational error before and after scaling}
\label{Tab:quan_scale2}
\begin{threeparttable}

\begin{tabular}{ |c|c|c|}
\hline
 & \makecell[c]{Rotational Error \\ (before scaling) } & \makecell[c]{Rotational Error \\ (after scaling) } \\
\hline
RMSE & ${0.6302}^{\circ}$ & ${0.6302}^{\circ}$ \\
\hline
Mean & ${0.4870}^{\circ}$ & ${0.4870}^{\circ}$ \\
\hline
Median & ${0.3896}^{\circ}$ &  ${0.3896}^{\circ}$ \\
\hline
Max & ${3.9706}^{\circ}$ &  ${3.9706}^{\circ}$ \\
\hline
S.D. & ${0.4183}^{\circ}$ &  ${0.4183}^{\circ}$ \\
\hline
\end{tabular}
\begin{tablenotes}
\footnotesize
\item S.D.: Standard Deviation
\end{tablenotes}
\end{threeparttable}
\end{table}

\subsection{Virtual Plane Computation}

\begin{figure}[tb]\centering            \includegraphics[width=\columnwidth]{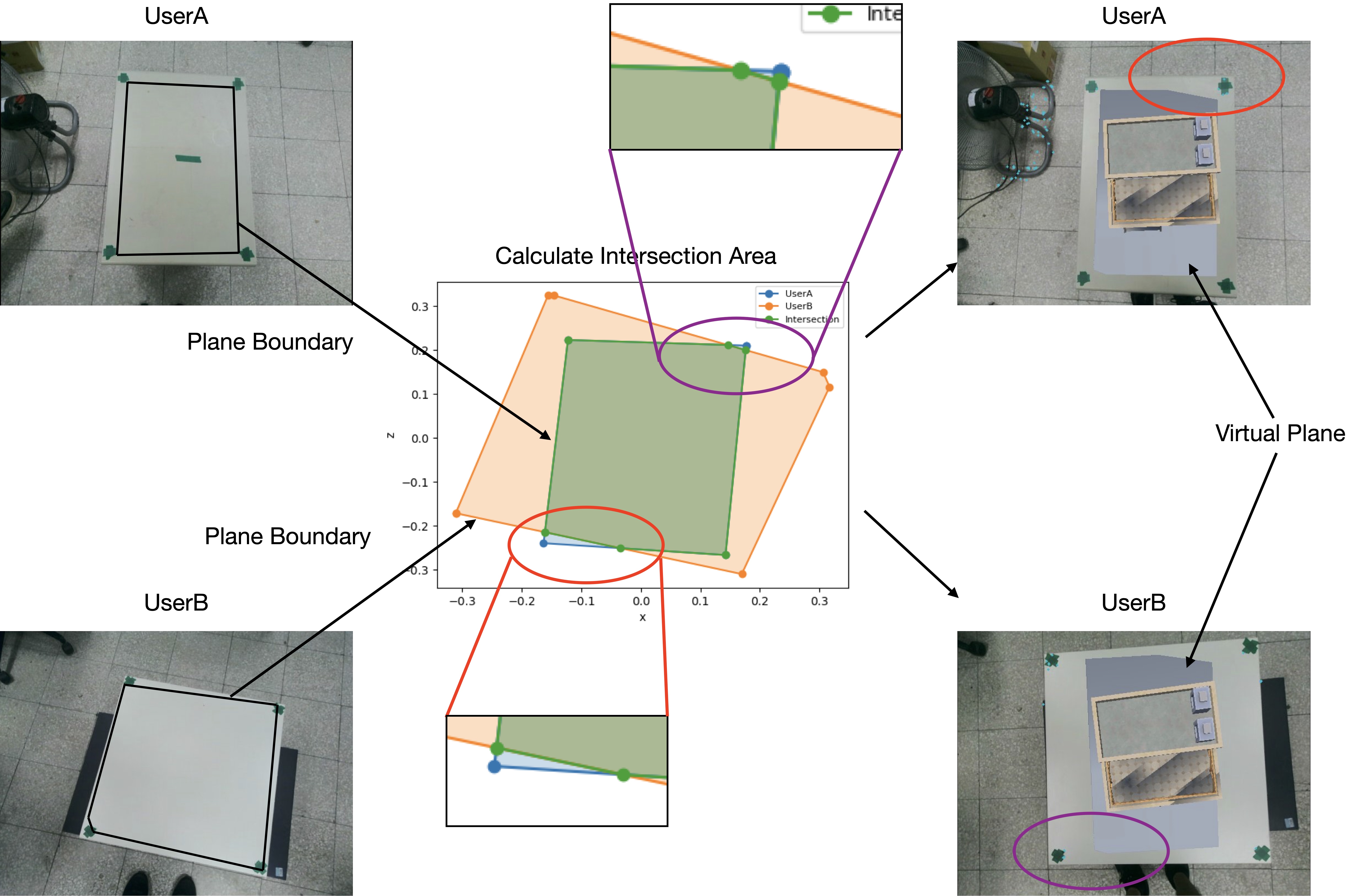}
            \caption{Visualization of Virtual plane computation in a Multi-User Remote Interactive Scenario.}\label{fig:intersection_exp}
\end{figure}

In previous section, we presented an algorithm for estimating the position and normal vector of a plane. Additionally, we identified the boundary points that encompass the estimated plane. The visualization of the virtual plane computation process is provided in Fig.~\ref{fig:intersection_exp}. In a multi-user remote interactive scenario, UserA and UserB may have tables of different sizes. After the plane estimation step, an estimated boundary is obtained to represent the surface area of each user's table. As shown in the left part of Fig.~\ref{fig:intersection_exp}, the black lines enclose the respective surface areas. Subsequently, both users transmit the boundary information to the coordination server, which calculates the intersection of the planes. The middle part of Fig.~\ref{fig:intersection_exp} illustrates the result of the intersection computation, and the intersection area is then sent back to the users. Finally, in the right part of Fig.~\ref{fig:intersection_exp}, the gray plane under the virtual house represents the intersection virtual plane. 

\subsection{Occlusion Rendering} 

\begin{figure}[tb]\centering            \includegraphics[width=\columnwidth]{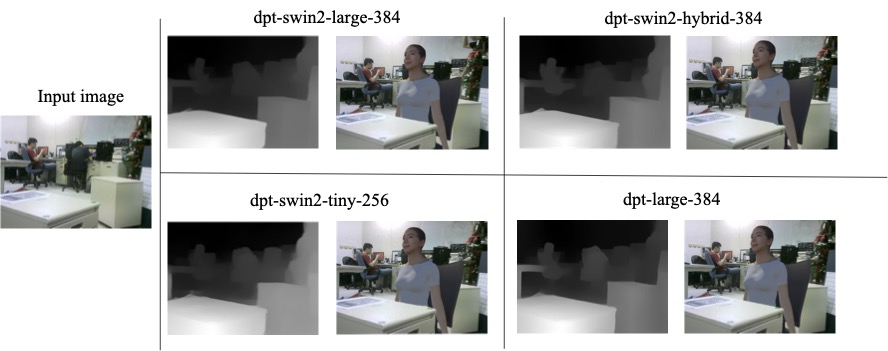}
            \caption{Depth maps and occlusion results of different models. The input image is taken from the side of the table.}\label{fig:occ_exp2}
\end{figure}

Our system integrates the monocular depth estimation model from MiDas\cite{b15}. MiDas provides several models with different network backbones and parameters, allowing us to balance performance and runtime based on the specific needs of augmented reality applications. In our experimental setup, we evaluated four specific models: dpt-swin2-large-384, dpt-swin2-tiny-256, dpt-hybrid-384, and dpt-large-384. The name "dpt" refers to their work on depth estimation called "Vision Transformers for Dense Prediction"\cite{b16}. In this work, they used different backbone architectures for their models. Specifically, "Swin2-large" and "Swin2-small" models utilize the Swin Transformer V2 as their backbones. On the other hand, the "hybrid" model combines a Vision Transformer (ViT) architecture \cite{b18} with ResNet-50. In this context, "Large" refers to the ViT Large variant. The numerical values associated with the model names represent the inference height of the input image. The runtime and the number of parameters for each model are presented in Table~\ref{Tab:runtime_occ}. Additionally, Fig.~\ref{fig:occ_exp2} show the depth maps generated by these four models and their corresponding occlusion results on a virtual avatar. The input image is taken from the side of the table in Fig.~\ref{fig:occ_exp2}.

\begin{table*}[tb]
    \caption{Runtime comparison of different models.}\centering
    \begin{tabular}{p{4cm} c c}
        \toprule[2pt]
         \textbf{Model} & \textbf{Number of parameters(M)} & \textbf{Runtime(fps)} \\
        \midrule[2pt] 
        dpt-swin2-large-384 & 213 & 10.5\\
        dpt-swin2-tiny-256 & 42 & 39.08\\
        dpt-hybrid-384 & 123 & 18.78 \\
        dpt-large-384 & 344 & 10.45 \\
        \bottomrule [2pt]
    \end{tabular}\label{Tab:runtime_occ}
\end{table*}

\begin{table*}[tb]
    \caption{Runtime performance evaluation.}\centering
    \begin{tabular}{c c c c}
        \toprule[2pt]
         \textbf{Thread} & \textbf{Module} & \textbf{Processing time(s)} & \textbf{Runtime(fps)} \\
        \midrule[2pt] 
        Main thread & ORB-SLAM2 & 0.067 & 15.004\\
        \midrule[2pt]
        \multirow{3}{*}{Depth estimation thread} & Depth estimation model & 0.053 & 18.779\\
        & Data transmission & 0.027 & - \\
        & Total & 0.0803 & 12.452 \\
        \midrule[2pt]
        \multirow{2}{*}{Coordination thread} & Two-users pose transmission & 0.055 & - \\
        & Three-users pose transmission & 0.07211 & - \\
        \bottomrule [2pt]
    \end{tabular}\label{Tab:runtime_performance}
\end{table*}
After comparing the runtime performance and visualization of the occlusion results, we observed that the dpt-swin2-tiny-256 model, while having a lower processing time, produces depth maps with unclear boundaries in the presence of complex objects. Consequently, the occlusion results are unsatisfactory. On the other hand, both the dpt-swin2-large-384 and dpt-large-384 models achieve excellent results. However, their processing times are significantly longer, making them impractical for real-time applications.

Considering these factors, we selected the dpt-hybrid-384 model as our depth estimation model for the experiments. It demonstrates a commendable balance between accuracy and processing speed. While it may not attain the perfection of the larger models, its occlusion results are satisfactory. This makes it a suitable choice for our requirements, as it provides a reasonable compromise between accuracy and real-time processing.

\subsection{Runtime Performance Evaluation}

We evaluate the overall runtime performance of our proposed multi-user positioning system. The results is shown in Table~\ref{Tab:runtime_performance}. In our system, we implement a three-threaded architecture to enable real-time localization and collaboration among users. The first thread, known as the main thread, executes the ORB-SLAM2 algorithm, which is responsible for user localization. The processing time of main thread is 0.067s per frame, equivalent to a frame rate of 15.004fps.

The second thread is responsible for depth estimation, encompassing both data transmission time and the actual depth estimation process. Data transmission, involving image transfer to the depth server and receiving the predicted depth map, takes 0.027s. Concurrently, the dpt-hybrid-384 depth estimation model runs for 0.053s, resulting in a total processing time of 0.0803s per frame at a frame rate of 12.452 fps. 

The third thread is the coordination thread. This thread is responsible for transmitting and receiving pose information among the users. By measuring the time interval between receiving pose information from a specific user, we evaluated the system's performance in two-users and three-users collaboration scenarios. For the two-users collaboration, the pose transmission time averaged at 0.055s, while in the three-users scenario, the average pose transmission time was 0.0721s.

We tested the runtime performance for these modules, however optimizing for communication delays are outside of the scope of this paper.

\section{Conclusion}
In this paper, we proposed a multi-user positioning system capable of accurately placing virtual objects on table surfaces and rendering them at correct angles using ORB-SLAM2\cite{b1}. To enable remote collaboration, we utilized the estimated plane as a reference frame to align the SLAM coordinate systems of multiple participants, thereby showcasing synchronized avatar movements in AR environments. Furthermore, we incorporated a monocular depth estimation model from Midas\cite{b15} to simulate occlusion effects. This system can serve as a tool for developers, offering essential technologies for the development of multi-user AR applications.

We conducted a series of experiments on the different modules within our system to validate their performance. The qualitative results of the scale calibration process and plane estimation module demonstrated the system's ability to accurately calculate the scale between the SLAM map and the real environment, and precisely place the virtual objects on the surface of the table. The quantitative and qualitative results of the positioning module show that our system can accurately track the user's pose and maintain the virtual object in the desired location. Our positioning module only has 0.0065 meters translational error and 0.6302 degrees rotational error in terms of RMSE compared with the commercial solution Vuforia. We also discussed the trade-off between runtime and performance in the depth estimation model with different backbone networks in augmented reality. To enable real-time applications, we selected a model that achieved satisfactory occlusion results and maintained an adequate frame rate. However, we recognize that our system doesn't scale well as the number of users increase. Hence, this should be explored in future research.
\acknowledgments{
This work was supported in part by a grant from Jorjin Technologies Inc. This research was also supported by the Ministry of Science and Technology of Taiwan, and Center for Artificial Intelligence Advanced Robotics, National Taiwan University, under the grant numbers MOST 110-2634-F-002-049 MOST 110-2221-E-002-166-MY3.}

\newpage
\bibliographystyle{abbrv-doi}

\bibliography{template}
\end{document}